\documentclass[fullpage,11pt]{article}
\usepackage[utf8]{inputenc}
\usepackage{url}
\usepackage{graphicx}
\usepackage{hyperref}
\hypersetup{
    colorlinks=true,
    linkcolor=blue,
    filecolor=magenta,      
    urlcolor=blue,
    citecolor=blue
}
\usepackage[margin=1in]{geometry}
\usepackage{setspace}
\usepackage{xspace}

\title{DBOS: A Proposal for a Data-Centric Operating System}
\author{The DBOS Committee\thanks{DBOS committee members in alphabetic order: \href{https://web.eecs.umich.edu/~michjc/}{Michael Cafarella} (MIT CSAIL), \href{http://pages.cs.wisc.edu/~dewitt/}{David DeWitt} (MIT CSAIL), \href{https://vijayg.mit.edu/}{Vijay Gadepally} (MIT LLSC), \href{http://www.mit.edu/~kepner/}{Jeremy Kepner} (MIT LLSC), \href{http://www.stanford.edu/~kozyraki}{Christos Kozyrakis} (Stanford University), \href{http://people.csail.mit.edu/kraska/}{Tim Kraska} (MIT CSAIL), \href{https://www.csail.mit.edu/person/michael-stonebraker}{Michael Stonebraker} (MIT CSAIL), and \href{https://cs.stanford.edu/~matei}{Matei Zaharia} (Stanford University).}\\ \\
{\small \href{mailto:dbos-project@googlegroups.com}{dbos-project@googlegroups.com}}
}
\date{}

\newcommand{\eg}{{\em e.g.}}
\newcommand{\dca}{DBOS\xspace}
\begin{document}

\maketitle

\abstract{
Current operating systems are complex systems that were designed before today's computing environments.
This makes it difficult for them to meet the scalability, heterogeneity, availability, and security challenges in current cloud and parallel computing environments.
To address these problems, we propose a radically new OS design based on \emph{data-centric architecture}: all operating system state should be represented uniformly as database tables, and operations on this state should be made via queries from otherwise stateless tasks.
This design makes it easy to scale and evolve the OS  without whole-system refactoring, inspect and debug system state, upgrade components without downtime, manage decisions using machine learning, and implement sophisticated security features.
We discuss how a database OS (DBOS) can improve the programmability and performance of many of today's most important applications and propose a plan for the development of a \dca proof of concept. 
}
\section{Introduction}
\label{sec:intro}

Current operating systems have evolved over the last forty years into complex overlapping code bases~\cite{tsai16,atlidakis16,lozi16,padioleau06}, which were architected for very different environments than exist today. The cloud has become a preferred platform, for both decision support and online serving applications. Serverless computing supports the concept of elastic provision of resources, which is very attractive in many environments.  Machine learning (ML) is causing many applications to be redesigned, and future operating systems must intimately support such applications. Hardware is becoming massively parallel and heterogeneous. These “sea changes” make it imperative to rethink the architecture of system software, which is the topic of this paper. 

Mainstream operating systems (OSs) date from the 1980s and were designed for the hardware platforms of 40 years ago, consisting of a single processor, limited main memory and a small set of runnable tasks.  Today’s cloud platforms contain hundreds of thousands of processors, heterogeneous computing resources (including CPUs, GPUs, FPGAs, TPUs, SmartNICs, and so on) and multiple levels of memory and storage. These platforms support millions of active users that access thousands of services. Hence, the OS must deal with a scale problem of $10^5$ or $10^6$ more resources to manage and schedule. Managing OS state is a much bigger problem than 40 years ago in terms of both throughput and latency, as thousands of services must communicate to respond in near real-time to a user's click~\cite{tail13, killer-usec}.

Forty years ago, there was little thought about parallelism.  After all, there was only one processor.  Now it is not unusual to run Map-Reduce or Apache Spark jobs with thousands of processes using millions of threads~\cite{LLMapReduce2016}.  Stragglers creating long-tails inevitably result from substantial parallelism and are the bane of modern systems: incredibly costly and nearly impossible to debug~\cite{tail13}. 

Forty years ago programmers typically wrote monolithic programs that ran to completion and exited.  Now, programs may be coded in multiple languages, make use of libraries of services (like search, communications, databases, ML, and others), and may run continuously with varying load.  As a result, debugging has become much more complex and involves a flow of control in multiple environments. Debugging such a network of tasks is a real challenge, not considered forty years ago.

Forty years ago there was little-to-no-thought about privacy and fraud.  Now, GDPR~\cite{gdpr} dictates system behavior for Personally Identifiable Information (PII) on systems that are under  continuous attack.  Future systems should build in support for such constructs.  Moreover, there are many cases of bad actors doctoring photos or videos, and there is no chain of provenance to automatically record and facilitate exposure of such activity.

Machine learning (ML) is quickly becoming central to all large software systems. However, ML is typically bolted onto the top of most systems as an after thought. Application and system developers struggle to identify the right data for ML analysis and to manage synchronization, ordering, freshness, privacy, provenance, and performance concerns. Future systems should directly support and enable AI applications and AI introspection, including first-order support for declarative semantics for AI operations on system data. 


In our opinion, {\it serverless computing} will become the dominant cloud architecture.  One does not need to spin up a virtual machine (VM), which will sit idle when there is no work to do.  Instead, one should use an execution environment like Amazon Lambda.  Lambda is an efficient task manager that encourages one to divide up a user task into a pipeline of several-to-many subtasks\footnote{In this paper, we will use Lambda as an exemplar of any resource allocation system that supports “pay only for what you use.”}.  Resources are allocated to a task when it is running, and no resources are consumed at other times.  In this way, there are no dedicated VMs; instead there is a collection of short-running subtasks.  As such, users only pay for the resources that they consume and their applications can scale to thousands of functions when needed.  We expect that Lambda will become the dominant cloud environment unless the cloud vendors radically modify their pricing algorithms.  Lambda will cause many more tasks to exist, creating a more expansive task management problem.

Lastly, “bloat” has wrecked havoc on elderly OSs, and the pathlength of common operations such as sending a message and reading bytes from a file are now uncompetitively expensive. One key reason for the bloat is the uncontrolled layering of abstractions. Having a clean, declarative way of capturing and operating on operating system state can help reduce that layering. 

These changed circumstances dictate that system software should be reconsidered. 
In this proposal, we explore a radically different design for operating systems that we believe will scale to support the performance, management and security challenges of modern computing workloads: a \emph{data-centric architecture} for operating systems built around clean separation of all state into database tables, and leveraging the extensive work in DBMS engine technology to provide scalability, high performance, ease of management and security.
We sketch why this design could eliminate many of the difficult software engineering challenges in current OSes and how it could aid important applications such as HPC and Internet service workloads.
In the next seven sections, we describe the main tenets of this data-centric architecture.  Then, in Section \ref{sec:proofofconcept}, we sketch a proposal concerning how to move forward.


\section{Data-Centric Architecture}
\label{sec:dca}

One of the main reasons that current operating systems are so hard to scale and secure is the lack of a single, centralized data model for OS state. For example, the Linux kernel contains dozens of different data structures to manage the different parts of the OS state, including a process table, scheduler, page cache, network packet queues, namespaces, filesystems, and many permissions tables. Moreover, each of the kernel components offers different interfaces for management, such as the dozens of APIs to monitor system state (/proc, perf, iostat, netstat, etc). This design means that any efforts to add capabilities to the system as a whole must be Herculean in scope. For example, there has been more than a decade of effort to make the Linux kernel more scalable on multicores by improving the scalability of one component at a time~\cite{boyd-wickizer-linux-2010,bhat-scaling-fs,linux-scaling-networking,lozi16}, which is still not complete. Likewise, it took years to add uniform security management interfaces to Linux -- AppArmor~\cite{apparmor} and SELinux~\cite{selinux} -- that have to be kept in sync with changes to the other kernel components. It similarly took years to enable DTrace~\cite{dtrace}, a heavily engineered and custom language for querying system state developed in Solaris, to run on other OSs. The OS research community has also proposed numerous extensions to add powerful capabilities to OSs, such as tracing facilities~\cite{feiner-dynamic-instrumentation,kedia-dynamic-instrumentation}, tools for undoing changes made by bad actors~\cite{retro}, and new security models~\cite{histar,song-data-flow-integrity}, but these remain academic prototypes due to the engineering cost of integrating them into a full OS.

To improve the scalability, security and operability of OSes, we propose a {\it data-centric architecture}: designing the OS to explicitly separate data from computation, and centralize all state in the OS into a uniform data model. In particular, we propose using database tables, a simple data model that has been used and optimized for decades, to represent OS state. With the data-centric approach, the process table, scheduler state, flow tables, permissions tables, etc all become database tables in the OS kernel, allowing the system to offer a uniform interface for querying this state. Moreover, the work to scale or modify OS behavior can now be shared among components. For example, if the OS components access their state via table queries, then instead of reimplementing dozens of data structures to make them scalable on multicores, it is enough to scale the implementations of common table operations. Likewise, new debugging or security features can be implemented against the tabular data model once, instead of requiring separate integration work with each OS component. Finally, making the OS state explicitly isolated also enables radical changes in OS functionality, such as support for zero-downtime updates~\cite{ksplice,kpatch}, distributed scale-out~\cite{legoos,multikernel}, rich monitoring~\cite{dtrace,google-perf-tracing}, and new security models~\cite{histar,song-data-flow-integrity}.

To manage the state in a data-centric operating system, we will require a scalable and reliable implementation of database tables. For this purpose, we simply recommend building the OS over a scale-out DBMS engine, leveraging the decades of engineering and operational experience running mission-critical applications. In other words, we suggest to build a {\it database operating system (\dca)}. While the DBMS engine will need some basic resource management functionality to bootstrap its execution, this could be done over a cluster of servers running current OSs, and eventually bootstrapped over the new \dca. Today, DBMS engines already manage the most critical information in some of the largest computer systems on the planet (e.g. cloud provider control planes). Thus, we believe that they can handle the challenges in a next-generation OS. Moreover, recent trends such as support for polystores~\cite{Tan2017,lu2018multi} that combine multiple storage engines will enable the DBMS to use appropriate storage strategies for each of the wide range of data types in an OS, from process tables all the way to file systems.

In more detail, this \dca approach results in several prescriptive suggestions as discussed in the next section.

\subsection{Prescriptive Suggestions}

\vspace{0.2cm}
\noindent {\bf All OS state should be stored in tables in the DBMS.}  Unix was developed with the mantra that {\it “everything is a file”}.  This mantra should be updated to {\it ``everything is a table''}, with first class support for high performance declarative semantics for query and AI operations on dense, sparse, and hypersparse tables~\cite{Graphulo2015, D4M2015, AssocArray2016, GraphBLAS2016, kepner2018mathematics, Cailliau_2019}.  For example, there should be a task table with the state of every task known to the system, a flow table with ongoing network flows, a set of tables to represent the file system, etc~\cite{kepner:2018}.

\vspace{0.2cm}
\noindent {\bf All changes to OS state should be through DBMS transactions.}  The OS will need to include multiple routines in complex imperative code to implement APIs or complex resource management logic, but when these routines need to access OS state, we will require them to do so through DBMS transactions. This choice offers several benefits. First, parallelism and concurrency become easier to reason about because there is a transaction manager to identify conflicts. Second, computation threads in the OS can safely fail without corrupting system state, enabling a wide range of features including geographic distribution, improved reliability, and hot-swapping OS code. Third, transactions provide a natural point to enforce security and integrity constraints as is standard in DBMSs today.

\vspace{0.2cm}
\noindent {\bf The DBMS should be leveraged to perform all functions of which it is capable.}  For example, files should be supported as blobs and tables in the DBMS.  As a result, file operations are simply queries or updates to the DBMS.  File protection should be implemented using DBMS security features such as view-based access controls for complex security policies.  In other words, there should only be ONE extensible security system, which will hopefully be better at avoiding configuration errors and leaks than the sprawl of configuration tools today.  Authentication should similarly be done only once using DBMS facilities. Finally, virtualization and containerization features can elegantly be implemented using database views: each container simply acts on a view of the OS state tables restricted to objects in that container.

As a result, ALL system data should reside in the DBMS. To achieve very high performance, the DBMS must leverage sophisticated caching and parallelization strategies and compile repetitive queries into machine code [2], as is being done by multiple SQL DBMSs, including Redshift [3].  A DBMS supports transactions, so ALL OS objects should be transactional.  As a result, transactions are implemented just once, and used by everybody.  

\vspace{0.2cm}
\noindent {\bf Decision support capabilities are facilitated.}  OSs currently perform many decision support and monitoring tasks. These include:

\begin{itemize}
    \item Choosing the next task to run
    \item Discovering stragglers in a parallel computation
    \item Finding over(under) loaded resources
    \item Discovering utilization for the various resources
    \item Predicting bottlenecks in real-time systems
\end{itemize}

All of these can be queries to the DBMS.  

\subsection{Tangible Benefits}

\vspace{0.2cm}
\noindent {\bf Performance optimization:}  OS kernel subsystems have often undergone extensive refactoring to improve performance by changing the data structures used to manage various state~\cite{lozi-wasted,linux-cfs,linux-refactoring-2017,linux-block-io-2016}. If the OS had been designed around a DBMS instead, many of these updates would amount to changing indexes or changing operator implementations in the DBMS (e.g., adding parallel versions of operators). Moreover, the DBMS approach would enable further methods to improve performance that are not implemented in OSes today, such as cost-based optimization (switching access paths for an operation based on the current data statistics and expected size of the operation) or adaptive mid-query reoptimization.

\vspace{0.2cm}
\noindent {\bf Security:} DBMS access control tools such as view, attribute and role based ACLs~\cite{systemr,attributeacl} can elegantly implement many of the security policies in SELinux, AppArmor and other OS security modules. Moreover, if these rules are implemented as view definitions or SQL statements within the DBMS, the security checking code can be compiled into the queries that regular OS operations run, instead of being isolated in a separate module that adds overhead to OS operations~\cite{selinux-overhead}.

\vspace{0.2cm}
\noindent {\bf Virtualization and containerization:} Tremendous engineering effort has gone into enabling virtualization and containerization in OSes over the past decade, i.e., enabling a single instance of the OS to host multiple applications that each get the abstraction of an isolated system environment. These changes have generally required modifying all data structures and a large amount of logic in the kernel to support different "namespaces" of objects for each container. With \dca, virtualization and containerization can elegantly be achieved using DBMS views: each container's DBMS queries only have access to a view that restricts to objects with that container ID, whereas a root user can have access to all objects. We believe that many queries and logic in OS components would not have had to be modified at all to add virtualization with this approach, other than being made to run on these views instead of on the raw OS state tables.

\vspace{0.2cm}
\noindent {\bf Geographic distributability:} After all, nodes in a cloud vendor’s offering are geographically distributed.  Transactional replication is a desired service of cloud offerings.  This can be trivially provided by a geographically dispersed DBMS.  This is in keeping with “implement any function only once; in the interest of simplicity”.

\vspace{0.2cm}
\noindent {\bf More sophisticated file management:}  Since files are stored in the DBMS, as blobs and tables, and the directory structure is a collection of tables, and SQL access control is used for protection, the large amount of code that implements current file systems, essentially disappears.  Also, we claim that current DBMSs which use aggressive compilation query and caching have gotten a great deal faster than the DBMSs of yesteryears.  Also, multinode main memory DBMSs such as VoltDB and MemSQL are capable of tens of millions of simple transactions per second.  Since a file read/write is just such a simple transaction, we believe that our proposed implementation can be performance competitive.  In addition, more sophisticated file search becomes trivial to implement.  For example, finding all files underneath a specific directory accessed in the last 24 hours that are more than 1GByte in size is merely a SQL query.  The net result is additional features, much less code and (hopefully) competitive performance.

\vspace{0.2cm}
\noindent {\bf Better scheduling:}  There will be task and resource tables in the DBMS capturing what tasks runs on cores, chips, nodes, and datacenters and what resources are available.  Scheduling thousands of parallel tasks in such environments as Map-Reduce and Spark is mainly an exercise in finding available resources and stragglers, because running time is the time of the slowest parallel task.  Finding outliers in a large task table is merely a decision support query that can be coded in SQL.  Again, we believe  that the additional functionality can be provided at a net savings in code.

\vspace{0.2cm}
\noindent {\bf Enhanced state management:} Using this approach it is straight-forward to divide application state into two portions.  The first is transient and can be stored in data structures external to the DBMS.  The second is persistent and must be stored in the DBMS transactionally.  Since replication will be provided for all DBMS objects, application failures can merely failover to a new instance.  This instance reads the persistent state from the DBMS and resumes the computation.  This failover architecture was pioneered by Tandem Computers in the 1980’s and can be provided nearly for free using our architecture.

Additional benefits accrue to this architecture by using a modern “server-less” application architecture, a topic which we defer to Section~\ref{sec:programmingmodel}.

\section{Task Communication}
\label{sec:communication}

Data communications can be readily expressed as operations on a geographically distributed DBMS.  A pull-based system can be supported by the sender writing a record into the DBMS and the receiver reading it.  A push-based system can be supported by the sender writing to the DBMS and setting a trigger to alert the receiver when he becomes active.  This can be readily extended to multiple senders and recipients.  In addition, DBMS transactions support exactly-once messages. Such an approach significantly simplifies programming allowing the programmer to easily implement non-blocking send programs that  have been demonstrated comparable bandwidth to more complex messaging systems~\cite{kepner2009parallel, byun2019}  

The CPU overhead of conventional TCP/IP communication is considered onerous by most, and new lighter-weight mechanisms, such as RDMA and kernel-bypass systems, are an order of magnitude faster~\cite{belay14,mitchell13}.  Hence, it seems reasonable to build special purpose lightweight communication systems whose only customer is the DBMS.   This has already been shown to accelerate DBMS transactions by an order of magnitude, relative to TCP/IP in a local area networking environment~\cite{zamanian19}, and it is possible that appropriate hardware could offer advantages of this approach in a wide area networking world.  As such, it is an interesting exercise to see if a competitive messaging system can be done through the DBMS.  It should also be noted that Amazon Lambda uses a storage-based communication system~\cite{lambda}.  Of course, a performant implementation would use something much faster than S3, such as a multi-node main memory DBMS.  

If this approach is successful, this will lower the complexity of future system software by replacing a heavyweight general purpose system with a lightweight and optimized, special purpose one.  It seems highly likely that the approach will work well in a hardware-assisted LAN environment.  WAN utilization seems more speculative.

\section{GDPR and Privacy Standards}
\label{sec:gdpr}

It is clear that privacy will be a future requirement of all system software.  GDPR~\cite{gdpr} is the European law that mandates ``the right to be forgotten''.  In other words, Personally Identifiable Information (PII) that a service holds on an individual must be permanently removed upon a user request.  In addition, data access must be based on the notion of ``purposes''.  Purposes are intended to capture the idea that performing aggregation for reporting purposes is a very different use case than performing targeted advertising based on PII data.  In SQL DBMSs access control is based on the notion of individuals and their roles.  These constructs have nothing to do with purposes, and a separate mechanism is required.  Obviously, this is a DBMS service. 

As noted in~\cite{storageForHealthcare}, a clean DBMS design can facilitate locating and deleting PII data inside the DBMS. However, one must also deal with the case where data is copied to an application and then sent to a second application.  Since all communication between applications goes through the DBMS, this message can be recorded by the DBMS,  allowing the DBMS to track PII data even when it goes out to applications. Of course, this will not prevent a malicious human from writing PII data to the screen and copying it outside of the system.  To deal with these kinds of leaks, applications must be ``sandboxed'' either virtually or cryptographically which can be readily incorporated into the database~\cite{Fuller2017,Yakoubov2014,popa2011cryptdb,CMD2014,poddar2019arx,CMD2015}. 

\section{Strong Provenance Guarantees}
\label{sec:provenance}

Data provenance is key to addressing many of the ills of modern data-centric life. Consider the following problems:

\vspace{0.2cm}
\noindent {\bf Data forging:} Detecting whether a photograph is doctored has become impossible for the typical news consumer. Even if a news service wants to provide trustworthy authorship information about its articles and photos, it has no trustworthy way to do so. Simply signing a photograph at the time it was taken is not sufficient, since there are some data-mutating operations (such as cropping or color adjustment) that news organizations must perform before publication.

\vspace{0.2cm}
\noindent {\bf Data debugging:}  Modern machine learning projects involve huge data pipelines, incorporating datasets and models from many different sources. Debugging pipeline output requires closely examining and testing these different inputs. Unfortunately, these inputs can come from partners with opaque engineering pipelines, or are incorporated in an entirely untracked manner, such as via a downloaded email attachment. As a result, simply enumerating the inputs to a data pipeline can be challenging, and fixing "root cause data problems" is frequently impossible.

\vspace{0.2cm}
\noindent {\bf Data spills:}  Today, an inadvertent data revelation is an irreversible mistake. There is no such thing as cleaning up after a database of social security numbers is mistakenly posted online. Although data handling practices must and can be improved, ensuring total data privacy today is a very difficult and brittle problem.

\vspace{0.2cm}
\noindent {\bf Data consumption and understanding:}  Much of modern life (as a professional, a consumer, and a citizen) consists of consuming and acting on data. The data processes that produce human-comprehensible outputs, such as the plots in a scientific article, are so complicated that it is quite easy for there to be errors that are undetectable even to the producer. Consider the case of economists Carmen Reinhart and Kenneth Rogoff, who in 2010 wrote an enormously influential article on public finance, cited by Representative Paul Ryan to defend a 2013 budget proposal, that was later found to be based on simplistic errors in an Excel spreadsheet~\cite{rogoff}. The authors did not acknowledge the error until three years after the paper was first written. Responsible data use means people must be able to quickly examine and understand the processes that yield the data artifacts all around us.

\vspace{0.2cm}
\noindent {\bf Data policy compliance:} Datasets and models often carry policies about how they can be used. For example, a predictive medical model might be appropriate for some age populations, but not others. Unfortunately, it is impossible for anyone, whether a data artifact producer or consumer, to have confidence about how data is being used.

\vspace{0.2cm}
A strong data provenance system would help address all of the above problems. All data operations by a modern operating system, such as copying, mutating, transmitting, etc., should be tracked and stored for possible later examination. It should be impossible to perform operations on a modern OS that sidestep responsible data provenance tracking.  Our proposed \dca architecture effectively logs all such operations, allowing an authoritative chain of provenance to be recorded.  (As with all the data the system collects, it will be stored in a DBMS.) This will support solutions to all of the above issues, requiring only log processing applications. Furthermore, first-class support for provenance throughout OS data structures will also simplify many system administration tasks, such as recovering from user errors or security breaches~\cite{chandra:2013}.

\section{Self-Adaptive via Modern ML}
\label{sec:adaptive}

Designing an operating system requires making assumptions about its future workload and data. These assumptions then materialize themselves as default parameters, heuristics, and various compromises. Unfortunately, all these decisions can significantly impact performance, especially if the assumptions turn out to be wrong. For example, if we assume that the OS mainly runs very short Lambda-like functions, then
reducing the overhead of starting a Lambda function may be more critical than optimal scheduling.  However, if we assume the workload is dominated by long-running memory intensive services, we require a very different scheduling algorithm, fair resource allocation strategies, and service migration techniques, whereas the startup time will matter very little. 
  
Moreover, operating systems offer a variety of knobs to tune the system for a particular workload or hardware. While providing flexibility, all the options put a burden on the administrator to set the knobs correctly and to adjust them in the case the workload, data, or hardware changes. 

To overcome those challenges, we suggest that \dca should be introspective, adaptable, and self-tuning through two design principles:

\vspace{0.2cm}
\noindent {\bf Knob-free design:} We believe that all parameters of the system should be designed to be self-tuning from the beginning. That is, \dca will deploy techniques similar to SmartChoices \cite{smartchoices} for all parameters and constants to make them automatically tuneable. The key challenge in globally optimizing all these parameters is then to gather and analyzing the state of the OS and the different components. Storing all this information in the OS database will significantly simplify the process and make true self-tuning possible. 

\vspace{0.2cm}
\noindent {\bf Learned components:} To address a wide range of use cases, the system developer often has to make algorithmic compromises. For instance, every operating system requires a scheduling algorithm, but the chosen scheduling algorithm might not be optimal under all workloads or hardware types. In order to provide the best performance, we envision that the system is able to automatically switch the algorithm used, based on the workload and data. This would apply to scheduling, memory management, etc~\cite{delimitrou13,cortez17}. 

\vspace{0.2cm}
In some cases it might be even possible to learn the entire component or parts of it. For example, recent results have shown that it is sometimes possible to learn a scheduling algorithm, which performs better than traditional more static heuristics \cite{DBLP:conf/sigcomm/MaoSVMA19,46646}. This learning of components would allow the system to more readily adapt to the workload and data, and perhaps provide unprecedented performance. 

To achieve a knob-free design and learned components, we suggest that the \dca needs to be designed from the beginning to be Reinforcement Learning (RL)-enabled. RL is the leading technique to tune knobs and build components based on the observed behaviour in an online fashion. Today, RL is usually added as an afterthought. This leads to several problems including difficulty in finding the right award function or supporting the required RL exploration phase.  In many cases this requires  the extra work of building a simulator or a light-weight execution environment to try out new approaches. By making RL a first-class citizen in the system design, we believe that we can overcome these challenges. Moreover, managing all state data in a database and making it analyzable, will again be a key enabler for this effort. 

If successful, the resulting system would be able to quickly adapt itself to changing conditions and provide unprecedented performance for a wide range of workloads while making the administration of the system considerably easier. 

\section{Diverse Heterogenous Hardware}
\label{sec:heterogeneity}

Managing compute, storage, and communication hardware is a primary function for an operating system. The key abstractions in existing operating systems were developed for the homogeneous hardware landscape of the last century. Kernel threads (processes), virtual memory, files, and sockets were sufficient to abstract and manage single-core computers with limited main memory backed by a slow hard disk, connected with low-bandwidth, high latency networking. 

Present-day hardware looks radically different. A single server machine contains tens to hundreds of cores in one or more chips, terabytes of main memory across a dozen channels, and multiple storage devices (SSDs and HDDs). The end of Dennard scaling~\cite{leiserson2020there} and the ascent of machine learning applications has led to the introduction of domain-specific accelerators like GPUs and TPUs, each with its own primitives for massively parallel computation and high-bandwidth memory~\cite{tpuv2}. The end of scaling for DRAM technology is motivating multi-level main memory systems using storage-class memories (SCM)~\cite{kamath2019storage}. Network interfaces allow direct access to remote memory at speeds faster than local storage. Beyond the single node, concepts such as multi-cloud, edge cloud, globally replicated clouds, and hardware disaggregation introduce heterogeneity in the type and scale of hardware resources. Existing operating systems were not designed for such scales or heterogeneity. This shortcoming is a primary culprit for the software bloat in applications and operating systems, including kernel bypass subsystems. Solutions have limited portability and are difficult to understand, debug, and reuse. 

Placing the operating system state in a DBMS introduces two properties that are useful in managing heterogeneous hardware. First, it clearly separates compute from data access. The operating system can manage data placement, caching, replication, and synchronization separately from the accelerated functions that operate on it. Second, it clearly separates control-plane from data-plane actions. One can improve or customize control-plane operations, such as scheduling, independently of the compute implementation using the best available accelerators.
 
To run efficiently on heterogeneous hardware,  \dca will be designed around two key principles.

\vspace{0.2cm}
\noindent {\bf Accelerated interfaces to DBMS:} \dca will implement the interfaces that allow heterogeneous hardware to interact with the DBMS, hiding the overall system scale and complexity. For example, the interface to a compute accelerator like a TPU can be a query that applies a user-defined function (UDF). The accelerator implements the UDF, while \dca implements the query that involves preparing inputs and outputs. This interface remains constant regardless if the accelerator is local, disaggregated, or in a remote datacenter. The accelerator state is stored in the DBMS to facilitate scheduling and introspection. \dca will directly manage memory and storage layers, as part of the DBMS resources available for data sharing, replication, or caching. \dca interfaces will leverage existing hardware mechanisms, such as virtual memory, as well as emerging mechanisms such as zero-copy/direct memory access networking interfaces or coherent fabrics (CXL). Over the time, hardware mechanisms will evolve to further accelerate the interactions between the DBMS and heterogeneous hardware. For example, SmartNICs will be optimized to accelerate DBMS interfaces, not just RDMA protocols, while GPUs and TPUs will directly support DBMS data operations.     

\vspace{0.2cm}
\noindent {\bf Accelerating the DBMS itself:} The performance and scalability of \dca itself relies heavily on the speed of DBMS operations. In addition to distributed execution and extensive caching, the DBMS will build upon modern hardware -- accelerators, storage class memory, and fast SmartNICs. Since all communication, dataplane, and control plane operations interface with the DBMS, the deployment of specialized accelerators for common DB operations like joins, filters, and aggregations will likely become essential~\cite{rapids}.

\section{Programming Model}
\label{sec:programmingmodel}

Historically, the programming model of choice was a single-threaded computation with execution interspersed with stalls for I/O or screen communication.  This model effectively requires multi-tasking to fill in for the stalls.  In turn, this requires interprocess protection and other complexity.  

Instead, we would recommend that everybody adopt the Lambda model, popularized by AWS~\cite{lambda}.  In other words, computation is done in highly parallel ``bursts,'' and resources are relinquished between periods of computation~\cite{gg19}.  This model allows one to give the CPU to one task at a time, eschewing multithreading and multiprogramming.  In addition, parallel processing can be done with a collection of short-lived, stateless tasks that communicate through the DBMS.The DBMS optimizes the communication by locally caching and co-scheduling communicating tasks when possible.  In effect, this is “server-less computing,” whereby one only pays for resources that are used and not for long-lived tasks.  Hence, under current cloud billing practices, this will save significant dollars.

That means \dca should adopt the Lambda model as well.  One should divide up a query plan into ``steps'' (operators).  Each operator is executed (in parallel) and then dies.  State is recorded in the DBMS.  Sharding of the data allows operator parallelism.  

Each Lambda task is given a exclusive set of resources, e.g., one or more cores until it dies.
 In the interest of simplicity and security, multi-tenancy and  multi-threading may be turned off.

There is a sharded scheduling table in the DBMS.  A task is runnable or waiting.  The scheduler picks a runnable task --- via a query --- and executes it.  When the task quits, the scheduler loops.  This will work well as long as applications utilize the Lambda model.

Dynamic optimization in the OS is gated by the time it takes stop, checkpoint, migrate, and restart applications/processes/threads.  In the cloud, this is often minutes, which means that very little dynamic optimization is possible.  Recent work has demonstrated that hand-coded fast launch (thousands of applications per second) is possible~\cite{Reuther_2018a, Reuther_2018b}.  This is all human-controlled static optimization~\cite{Byun_2019}.  The optimizing scheduler in \dca should be able to do this dynamically and launch millions of applications per second~\cite{kepner:2018}.

\section{Plans for a Proof of Concept}
\label{sec:proofofconcept}

Obviously, \dca is a huge undertaking. An actual commercial implementation will take tens of person-years.  As such, we need to quickly validate the ideas in this document.  Hence, we discuss demonstrating the validity of the ideas and then discuss convincing the systems community that \dca is worth the effort involved.

\begin{figure}
    \centering
    \includegraphics[width=\linewidth]{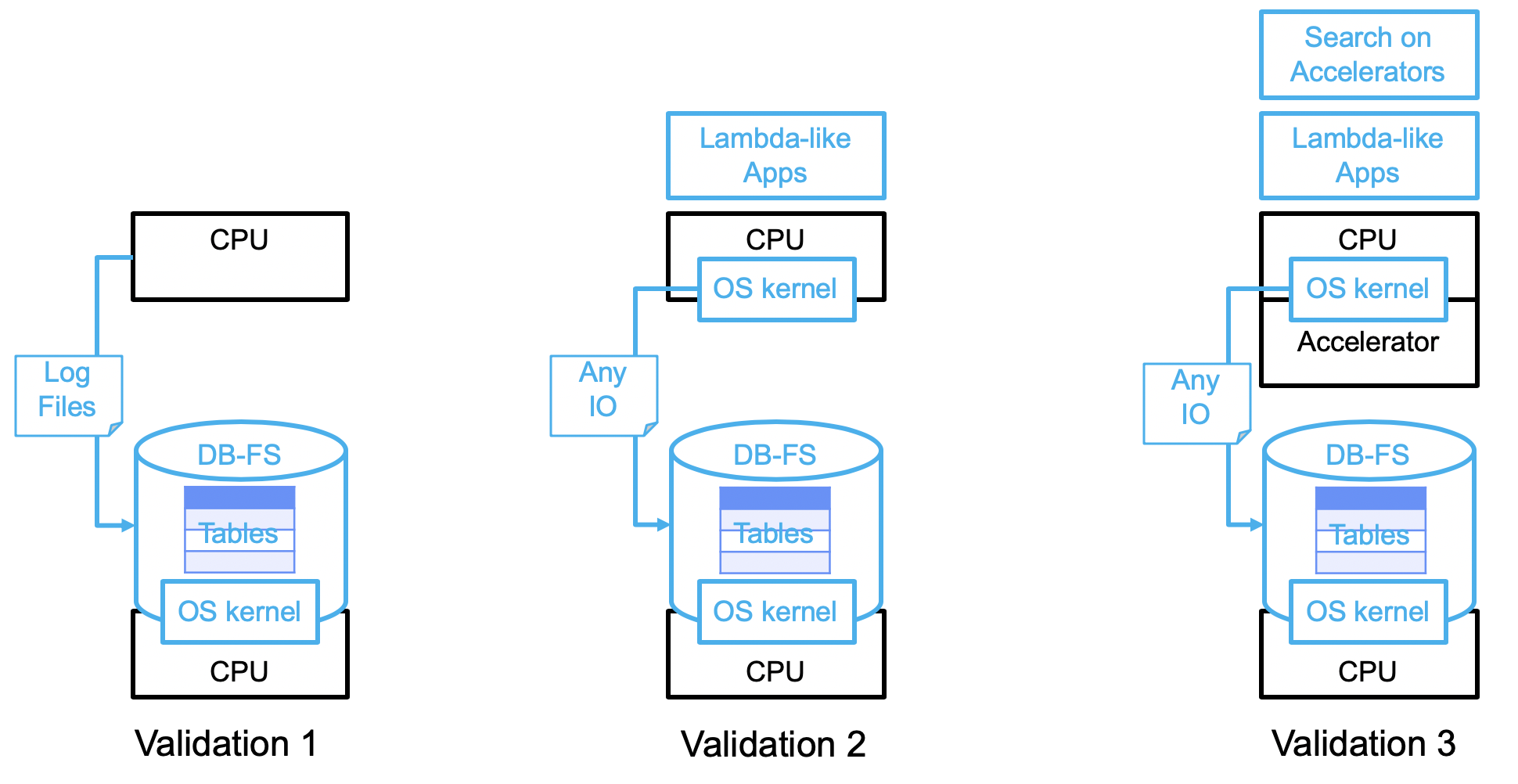}
    \caption{The three stages of our intended proof of concept. Phase 1 comprises a proof of concept to demonstrate database performance. Phase 2 includes work necessary for log processing. Phase 3 shows how we can manage accelerators and implement end-to-end microservice applications.}
    \label{fig:proofofconcept}
\end{figure}

\subsection{Key Characteristics}

A key challenge is to show a DBMS capable of acceptable performance and scalability to form the foundation of \dca. We believe that such a system should have the following characteristics:

\vspace{0.2cm}
\noindent {\bf Multi-core, multi-node executor:}  Many DBMSs support this today.

\vspace{0.2cm}
\noindent {\bf Server-less architecture:} Commercial DBMSs are moving toward allocating CPU resources on a per-query basis.  Snowflake has moved aggressively in this direction, based on a distributed file system (S3) aggressive caching and sharding only for CPU resources~\cite{snowflake}.

\vspace{0.2cm}
\noindent {\bf Polystore architecture:} Clearly, \dca will need to manage data from heterogenous sources such as process tables, schedulers, network tables, namespaces, and many permissions tables. It is likely that any single data management system will be able to efficiently manage the diversity and scale of the associated data structures. Different OS functionality will naturally fit into different types of storage engines and a polystore architecture~\cite{Tan2017,lu2018multi} can provide a single interface to these disparate and federated systems. A critical system characteristic would be to avoid developing a ``one size fits all'' \cite{stonebraker2018one, BigDAWG2016a, Khan2019} solution that is incapable of adapting as new types of data are collected and managed by \dca.

\vspace{0.2cm}
\noindent {\bf Open source code:} Obviously any code in a \dca prototype should be readily available.

\vspace{0.2cm}
\noindent {\bf Lambda-style, serverless runtime system:}  This will facilitate optimizing resource allocation.

Possible choices include SciDB, Presto, Accumulo, etc. We think the best option is to start with a prototype that comprises a DBMS built on an MIT Lambda-style system.

We view the key design choices of AWS Lambda as reservation-free, fixed-resource service for short-lived functions and will embody those in our own system. Other choices in today's commercial version of Lambda, such as S3 as the exclusive storage system, or the lack of direct communication between functions, seem like they should be rethought. It is unclear whether uniform resource constraints on the Lambda functions is a key design choice, or whether the system should offer heterogeneous resource constraints to enable a  more flexible development environment. 

We would expect to replace S3 as the storage system with something much faster~\cite{reflex17,pocket}, based on the discussion earlier.  We expect in one person year, we could demonstrate a LAN-based system along these lines.  We would then expect to test the performance of this prototype in two contexts.  The first goal is to provide file system performance comparable to today’s systems.  In addition, we expect to show our communication implementation can be comparable or faster to traditional TCP/IP networking.

To bootstrap running the DBMS itself, we plan to rely on minimal operating systems that have already been designed for cloud environments, such as unikernels, Dune or IX~\cite{dune12,belay14}, which are designed to run one application at a time and to give it high-performance access to the hardware. We will also make sure that the DBMS runs on Linux systems for easy development. The main facilities that the DBMS needs to bootstrap are a boot and configuration process, network access (which can also be used for logging), threads, and an interface to access storage. In the latter case, because the DBMS will manage all large data structures, raw block access may be sufficient. Today’s minimal OSes already support these facilities for hosting server applications as efficiently as possible in virtualized datacenters.

\subsection{Demonstration of Utility: Log Processing}
\label{sec:logs}
As a first example of using \dca to improve current OS functionality, we will implement a data-centric log processing and monitoring infrastructure in \dca that can monitor applications using existing OSes such as Linux. OSes, Networks, Schedulers, and File Systems generate enormous amounts of logs and metadata which are mostly kept in raw files.  Attempts to put these in databases (OS logs to Splunk; Network logs to NetAPP; Scheduler logs to MySQL; File System metadata to MySQL) barely meet minimal auditing requirements.  

A DBMS-based OS that organically stored these data in a high-performance database with first class support for dense, sparse, and hypersparse tables would be a huge win as it would make these data readily analyzable and actionable. It would also be able to execute streaming queries to compute complicated monitoring views in real time in order so simplify system management; simple metrics such as “how many files has each user created” can sometimes take hours to run with today’s file systems and OSes. Our team has conducted experiments showing the high-performance databases such as Apache Accumulo,  SciDB, and RedisGraph can easily absorb this data while enabling analysis that are not currently possible~\cite{kepner:2018, Cailliau_2019, Kepner_2019, kepner202075000000000}.  For example, "All files touched by a user during a time window", "Largest 10 folders owned by a user", "Computing cycles consumed by an application during a time window",  "Network traffic caused by a specific application", ...  These are very important questions for Cloud operators and very difficult to answer and require custom built tools to do so.  A DBMS OS should be able to answer these questions by design.

\subsection{Demonstration of Utility: Managing Accelerators}
\label{sec:accelerators}
In Section~\ref{sec:heterogeneity}, we discussed DBMS support for heterogeneous hardware, GPUs and FPGAs, based  on user-defined DBMS functions.  Our plan is to implement a prototype of this functionality to demonstrate its feasibility and performance. 

One of the defining features of modern datacenters is hardware heterogeneity. Far from being a uniform pool of machines, datacenters offer machines with different memory, storage, processing, and other capacities. Most notably, different machines offer vastly different accelerator capacities. Although GPUs for machine learning tasks comprise the most common class of accelerator, datacenters also contain FPGAs and other accelerators for video processing and encryption applications. These accelerators can be expensive: it is not feasible to outfit every machine in a large system with a top-flight GPU. Matching a heterogeneous workload to a heterogeneous pool of resources is a complicated and important task that is tailor-made for machine- rather than human-driven optimization. 

To address this challenge, we need to first design the DBMS-based API in \dca allows for portability. The same user code can drive execution on a local or remote GPU. Next, we need to exploit the flexibility of Lambda-style task allocation and the visibility into system state through the DBMS in order implement scheduling algorithms that utilize better the datacenter resources that  naive server-centric allocation schemes. We will demonstrate this functionality using by running a range of workloads on small clusters and by simulating larger, datacenter environments.


\subsection{Demonstration of Utility: End-to-end Microservice Applications}

Since \dca is designed around a distributed DBMS, it is a natural fit for data mining applications like the log processing discussed in Section~\ref{sec:logs}. However, it is not as obvious a match for online-serving applications, such as social networks, e-commerce sites, and media services, that consume large fractions on cloud systems. These applications consist of tens to thousands of microservices that must quickly communicate and respond to user actions within tight service level objectives (SLOs)~\cite{deathstar19}. Some microservices are simple tasks, such as looking up session information, while others are complicated functions such as recommendation systems based on neural networks or search functions using distributed indices. Microservice applications form the bulk of software-as-a-service products today and are the most critical operational applications for many organizations.

We will prototype an end-to-end microservices workload, such as a Twitter-like social network, in order to evaluate \dca's feasibility for these applications. During this process, we will answer two key questions. First, can \dca support the computation and communication patterns of such latency-critical applications in a performant manner? Second, can \dca help address the challenges in developing, scaling, and evolving such applications over time? 

With \dca, a social network will be implemented as a collection of serverless functions operating on multiple database tables. This presents multiple opportunities for performance optimization. For example, \dca can colocate communication functions to avoid remote communication, or selectively introduce new caching layers and indexes. Accelerators are also now used in many components of microservice applications, such as recommendation engines for social network content and search result re-ranking, so we will use the accelerator management capabilities in Section~\ref{sec:accelerators} to automatically offload and optimize these tasks.

Finally, because \dca uses a serverless model, data management decisions such as sharding and replicating datasets or evolving schemas are separated from the application code. This makes it significantly easier for application developers to implement architectural changes that are very difficult in microservice applications today. We will show how to use \dca to easily implement several such architectural changes:
\begin{enumerate}
    \item Changing the partitioning and schema of data in the application to improve performance (a common type of change that requires large engineering efforts in today’s services).
    \item Changing the partitioning of compute logic, e.g., moving from a “monolith” of co-located functions to separately scaling instances for different parts of the application logic.
    \item Making the application GDPR-compliant, by storing each user’s data in their geographic region and using the data provenance features of \dca to track which data was derived from each user or delete it on-demand.
    \item Changing the security model (\eg, which users can see data from minors or from European citizens) without having to refactor the majority of application code.
\end{enumerate}

\section{Conclusions}
\label{sec:conclusion}

We have presented a dramatically simpler view of systems software which avoids implementing the same functions in multiple components.  Instead, the architecture bets on a sophisticated DBMS to implement most functionality.  Section~\ref{sec:proofofconcept} suggested initial experiments to demonstrate feasibility.  Obviously, these steps should be carried out first.  

Following that, there are still many unanswered questions.  The most notable one is “Can this scale to a million nodes?”.  To the best of our knowledge, nobody has built a distributed DBMS at this scale.  Clearly, there will be unforeseen bottlenecks and inefficiencies to contend with.  Managing storage for a 1M node DBMS will be a challenge.  
A second question is “Can this be built to function efficiently?”  Section~\ref{sec:proofofconcept} discussed the file system and IPC.  However, memory management, caching, scheduling, and outlier processing are still issues. Obviously, the next step is to build a full-function prototype to answer these questions.

\newpage
\setstretch{1.0}
\bibliography{main}
\bibliographystyle{abbrv}

\end{document}